\documentclass{epl}

\title{Test of the isotopic and velocity selectivity of a lithium atom interferometer by magnetic dephasing}
\shorttitle{Isotopic and velocity selectivity of an atom
interferometer}
\author{M. Jacquey\inst{1} \and A. Miffre\inst{1,2} \and M. B\"uchner\inst{1}
\and G. Tr\'enec\inst{1} \and J. Vigu\'e\inst{1}
\thanks{E-mail:\email{jacques.vigue@irsamc.ups-tlse.fr}}}

\institute{ \inst{1} Laboratoire Collisions Agr\'egats
R\'eactivit\'e -IRSAMC
\\Universit\'e Paul Sabatier and CNRS UMR 5589
\\118, Route de Narbonne; 31062 Toulouse Cedex, France\\
\inst{2} LASIM, Universit\'e Claude Bernard Lyon I and CNRS UMR
5579
\\ 10, Rue A.M. Ampère; 69622 Villeurbanne Cedex, France}

\date{\today}
\begin{document}

\maketitle

\begin{abstract}

A magnetic field gradient applied to an atom interferometer
induces a $M$-dependent phase shift which results in a series of
decays and revivals of the fringe visibility. Using our lithium
atom interferometer based on Bragg laser diffraction, we have
measured the fringe visibility as a function of the applied
gradient. We have thus tested the isotopic selectivity of the
interferometer, the velocity selective character of Bragg
diffraction for different diffraction orders as well as the effect
of optical pumping of the incoming atoms. All these observations
are qualitatively understood but a quantitative analysis requires
a complete model of the interferometer.

\end{abstract}

If an inhomogeneous magnetic field is applied on a matter wave
interferometer, the phase of the interference pattern is modified,
provided that the matter wave has a non-zero magnetic moment. This
type of situation was first considered
\cite{aharonovo67,bernstein67} as a test of the sign reversal of a
spin $1/2$ wave function by a $2\pi$ rotation. This effect was
predicted since the foundation of quantum mechanics but considered
for a long time as not observable. The first successful
experimental test was made by H. Rauch and co-workers
\cite{rauch75} in 1975 with their perfect crystal neutron
interferometer and this work has been followed by several other
experiments reviewed in the book of Rauch and Werner
\cite{rauch00}.

Similar experiments can be done by applying a magnetic field
gradient on an atom interferometer: the fringe patterns
corresponding to the various Zeeman sub-levels experience
different phase-shifts and, when the gradient increases, the
fringe visibility exhibits a series of minima and recurrences, as
first observed by D. Pritchard and co-workers
\cite{schmiedmayer94,schmiedmayer97} and by Siu Au Lee and
co-workers \cite{giltner}. In this letter, we use our lithium atom
interferometer to show that the dependence of the fringe
visibility with the applied gradient gives a direct test of the
selective character of our interferometer with respect to the atom
velocity, to its isotopic nature and to its internal state
distribution. The velocity selective character of our atom
interferometer \cite{delhuille02a,miffre05b} comes from the use of
Bragg diffraction on laser standing waves. The choice of the laser
wavelength gives access to the isotopic selectivity of the
interferometer. Finally, by optical pumping $^7$Li in its $F=1$
ground state, we observe the effect of the internal state
distribution on the visibility variations.

\begin{figure}[t]
  \begin{center}
  \includegraphics[width = 7 cm,height= 5 cm]{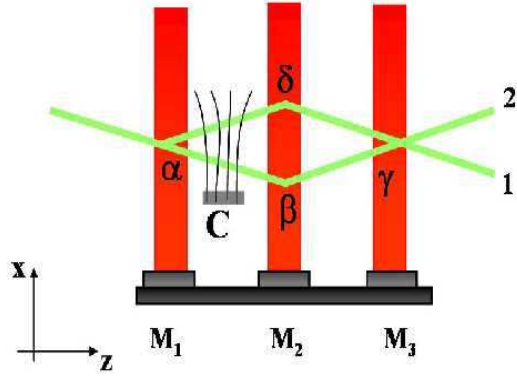}
\caption{\label{setup} Schematic drawing of our Mach-Zehnder atom
interferometer: a collimated atomic beam is diffracted by three
laser standing waves, produced by reflecting three laser beams on
three mirrors $M_i$. The output beams labelled $1$ and $2$ are
complementary, one of them (usually beam $1$) being detected. A
coil $C$ close to the atomic beams creates a magnetic field
gradient in the $\mathbf{x}$-direction.}
  \end{center}
  \end{figure}

\section{Calculation of the magnetic dephasing effect}

A Mach-Zehnder atom interferometer, as represented in figure
\ref{setup}, is operated with a paramagnetic atom. If the magnetic
field direction varies slowly enough, no spin flip occurs during
the atom propagation and the projection $M_F$ of the total angular
momentum $\mathbf{F}$ remains a good quantum number, the
quantization axis being parallel to the local magnetic field. In
the presence of a transverse gradient of the magnetic field, the
Zeeman energy $\Delta E\left( F, M_F\right)$ of the $F, M_F$
sub-level is not the same on the two atomic paths and, in the
perturbative limit (Zeeman energy considerably smaller than the
atom kinetic energy $\hbar^2 k^2/2m$), this energy difference
induces a phase shift equal to:

\begin{equation}
\label{n1}  \Delta \phi \left(F,M_F\right) =- \frac{1}{\hbar
v}\oint \Delta E\left(F, M_F, s \right) ds
\end{equation}

\noindent where the path integral follows the
$\alpha\beta\gamma\delta\alpha$ circuit (see figure \ref{setup})
and $v$ is the atom velocity. The interferometer signal is the
incoherent sum of the signals due to the various $F, M_F$
sub-levels:

\begin{eqnarray}
\label{n2} I &=& \sum_{F,M_F} \int dv I\left( F, M_F, v \right)
\nonumber \\ I\left( F, M_F, v \right)  &=&  I_0 P\left( v \right)
P\left( F, M_F \right) \times [1 + {\mathcal V}_0 \cos\left( \psi
+ \Delta \phi \left(F,M_F\right) \right]
\end{eqnarray}
\noindent $I\left( F, M_F, v \right)$ is the contribution of the
$F, M_F$ atoms with the velocity $v$. $P\left( F, M_F \right)$ and
$P(v)$ represent the internal state and velocity distribution of
the output flux. The fringe visibility ${\mathcal V}_0 $ is
assumed to be independent of the sub-level. Finally, the origin of
phase $\psi$ is explained below. We simplify the present
discussion by assuming that the Zeeman energy $ \Delta E\left( F,
M_F\right)$ is a linear function of the field $B$:

\begin{equation} \label{n3}  \Delta E\left(F,M_F\right)
 = - g_F \mu_B M_F B
\end{equation}
\noindent but our calculations take into account the non-linear
Zeeman terms due to hyperfine uncoupling which are non-negligible,
especially for $^6$Li. $\mu_B$ is the Bohr magneton and $g_F$ is
the Land\'e factor equal to $g_F = + 2/3$ (resp. $-2/3$) for the
$F=3/2$ (resp. $F=1/2$) level of $^6$Li and $g_F = + 1/2$ (resp.
$- 1/2$) for the $F=2$ (resp. $F=1$) level of $^7$Li, where the
nuclear magnetic moments have been neglected. The phase shift
$\Delta \phi \left(F,M_F\right)$ is given by:

\begin{equation}
\label{n4}  \Delta \phi \left(F,M_F\right) = \frac{g_F \mu_B M_F
}{\hbar v} \int_{z_{\alpha}}^{z_{\gamma}}
\frac{\partial\left|B(z)\right|}{\partial x} \Delta x(z) dz
\end{equation}
\noindent where $\Delta x(z)$ is the distance between the two
atomic beams in the interferometer and the integral is taken along
a path at mid-distance between the two paths $\alpha\beta\gamma$
and $\alpha\delta\gamma$ followed by each atom in the
interferometer.

As the coil used to create the magnetic field is small, the
magnetic gradient is important in a region where the field due to
the coil is substantially larger than the ambient field, which can
be neglected in the calculation. We have verified that this
approximation is good. The phase shift is then proportional to the
coil current $\mathcal{I}$ and to $v^{-2}$. One $v$ factor,
apparent in equation (\ref{n1}), comes from the time spent in the
perturbation. The other $v$ factor comes from the distance $\Delta
x(z)$, proportional to the diffraction angle $\theta_{diff} = 2p
h/(mva)$, where $p$ is the diffraction order and $a$ the grating
period. We thus get:

\begin{equation} \label{n5}
\Delta \phi \left(F,M_F\right) = C \frac{p g_F M_F \mathcal{I} }{m
v^2}
\end{equation}
\noindent where $C$ gathers several constant factors. It is
interesting to note that the equations (\ref{n1}-\ref{n5}) are
valid for bosons as well as for fermions. In the introduction, we
have recalled the discussion of the $4 \pi$ symmetry of fermions
\cite{aharonovo67,bernstein67} and the fact that our equations
take the same form for bosons and fermions may seem in
contradiction with well known results. The explanation of this
apparent contradiction lies in the fact that the phase shift
$\Delta \phi \left(F,M_F\right)$ is the product of a rotation
angle by the $M_F$ value. For fermions, $M_F$ is an half-integer
and the rotation angle must be equal to a multiple of $4 \pi$ for
a revival while the rotation angle must only be a multiple of
$2\pi$ for bosons.

We assume that the velocity distribution is given by:

\begin{equation} \label{n6} P(v)  =
\frac{S_{\|}}{u \sqrt{\pi}} \exp\left[-\left(
(v-u)S_{\|}/u\right)^2\right]
\end{equation}

\noindent where $u$ is the most probable velocity and $S_{\|}$ the
parallel speed ratio. This formula is used for supersonic beams
\cite{haberland85} but we have omitted a $v^3$ pre-factor, which
has minor effects when $S_{\|}^2$ is large , which is the case
here. The parallel speed ratio $S_{\|}$ can be varied by changing
the pressure in the supersonic beam source or the nozzle diameter
and the velocity distribution can be directly measured thanks to
Doppler effect by laser induced fluorescence of the lithium beam
\cite{miffre04,miffre05}.

Moreover, in the present calculations, $P(v)$ describes in fact
the product of the initial beam velocity distribution $P_i(v)$ by
the transmission $T(v)$ of the atom interferometer. Our
calculations show that the transmission $T(v)$ is roughly a
Gaussian function of the velocity around a velocity corresponding
to the Bragg condition.

\section{Some experimental details}

\begin{figure}[t]
\begin{center}
\includegraphics[width = 7 cm,height= 5 cm]{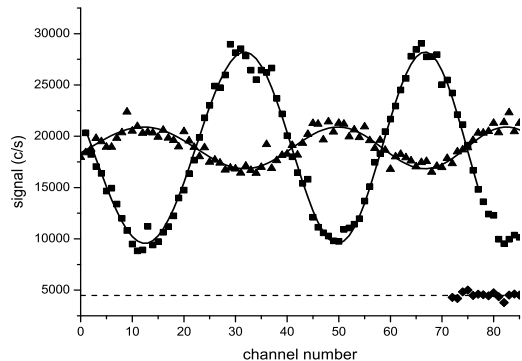}
\caption{\label{fringes} Interference fringes recorded
corresponding to different currents, $\mathcal{I}=0$ A (squares)
and $\mathcal{I} = 1.4 $ A (triangles), and their fits (full
curves). The interferometer was tuned for $^7$Li with first order
diffraction. The phase shift between the two experiments is very
close to $\pi$, corresponding to a visibility inversion. Each data
point corresponds to a $0.1$ s counting time. A few isolated data
points, due to bursts of the hot-wire detector, are not included
in the fits. The dotted line gives the mean value of the detector
background, recorded by flagging the beam.}
\end{center}
\end{figure}

Our atom interferometer \cite{delhuille02a,miffre05b} is a three
grating Mach-Zehnder interferometer. We use a supersonic beam of
argon seeded with natural lithium ($92.4$\% of $^7$Li and $7.6$\%
of $^6$Li). In the absence of optical pumping, the lithium atoms
are equally distributed over the $F,M_F$ hyperfine sub-levels of
their $2S_{1/2}$ ground state. The lithium mean velocity $u$ is $u
\approx 1065$ m/s. The gratings being laser standing waves, their
period $a$ is equal to half the laser wavelength $\lambda_L
\approx 671$ nm, chosen very close to the first resonance line of
lithium. We do not reiterate here the laser beam parameters which
are given in the full description of our interferometer
\cite{miffre05b}. The phase of the interference fringes depends on
the $x$-position of the gratings depending themselves on the
position $x_i$ of the mirrors $M_i$ forming the three laser
standing waves: this is the origin of the phase term $\psi$ in
equation (\ref{n2}), $\psi = 2p k_L (x_1+x_3-2x_2)$, where $k_L =
2 \pi/\lambda_L$ is the laser wavevector and $p$ is the
diffraction order. Figure \ref{fringes} shows experimental
interference fringes, observed by scanning the position $x_3$ of
mirror $M_3$ (this is the usual way of observing fringes in atom
interferometers as this phase is independent of atom velocity).

The magnetic field gradient is produced by a $3$ cm diameter coil,
with its axis at $4$ cm before the second laser standing wave. On
the coil axis, the distance $\Delta x$ between the two atomic
beams is about $94$ $\mu$m. The ambient field is roughly equal to
the Earth magnetic field with a $\sim 4 \times 10^{-5}$ Tesla
vertical component and a smaller horizontal component. From the
coil dimensions, we can calculate the magnetic field and its
gradient everywhere, but the distance of the coil to the atomic
beams, about $0.7$ cm, is not accurately known and we will
consider the constant $C$ appearing in equation \ref{n5} as an
adjustable parameter. With our maximum current $\mathcal{I}= 9 $
A, the maximum field seen by the atoms is $ B \approx 1.3 \times
10^{-3}$ T, sufficient to introduce some hyperfine uncoupling,
especially for $^6$Li isotope. As already stated, this effect is
taken into account in our calculations

During an experiment, we first optimize the interferometer fringes
with a vanishing coil current $\mathcal{I} = 0$, then we record a
series of interference signals as in figure \ref{fringes}, with
increasing values of $\mathcal{I}$. Slow drifts of the fringe
phase and visibility are corrected by frequent recordings with
$\mathcal{I} = 0$. From each recording, we can extract the phase
and the visibility of the interference pattern, from which we
deduce the effects of the applied field gradient, namely the
relative visibility ${\mathcal{V}}_r( \mathcal{I})= {\mathcal{V}}(
\mathcal{I} )/ {\mathcal{V}}(\mathcal{I}=0)$ and also the phase
shift $\Delta\phi(\mathcal{I})$ which will be discussed in another
paper.

\section{Test of the isotopic selectivity}

\begin{figure}[t]
\begin{center}
\begin{tabular}{cc}
\includegraphics[width =7 cm,height= 5 cm]{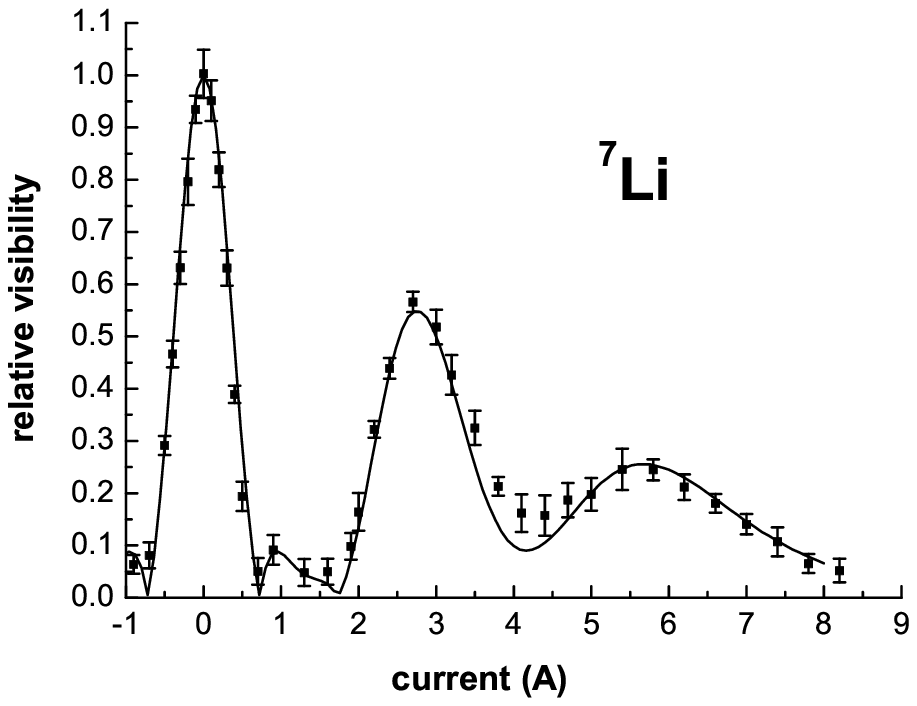} &\includegraphics[width =7cm,height= 5 cm]{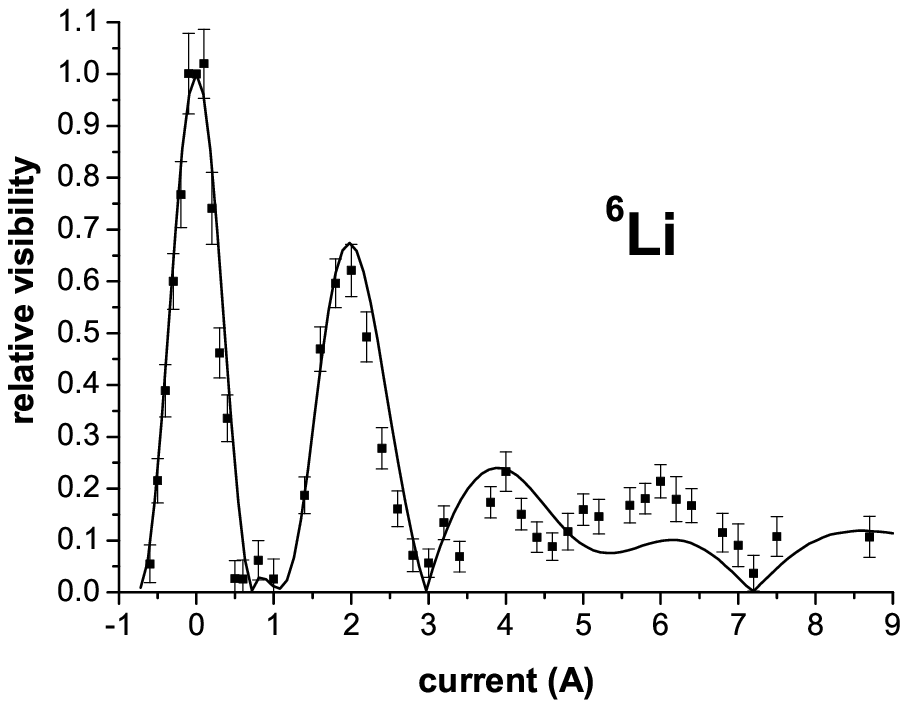}
\end{tabular}
\caption{\label{visibility6and7} Relative visibility
${\mathcal{V}}_r$ as a function of the applied current for $^7$Li
(left panel) and $^6$Li (right panel). Experimental data points
are represented by dots and the fits by full curves.}
\end{center}
\end{figure}

Here, we compare two experiments involving the two isotopes of
lithium and using first order diffraction $p=1$. We tune the
interferometer by choosing the laser wavelength, for $^7$Li on the
blue side (at $3$ GHz) of $^2$S$_{1/2}$ - $^2$P$_{3/2}$ transition
of $^7$Li and, for $^6$Li, on the red side (at $2$ GHz) of
$^2$S$_{1/2}$ - $^2$P$_{3/2}$ transition of $^6$Li. The nearest
transition of the other isotope is detuned from the laser by $14$
GHz in the first case and $12$ GHz in the second case. The
relative visibility is plotted as a function of the current
$\mathcal{I}$ in figure \ref{visibility6and7} for both isotopes.
The $\mathcal{I}=0$ visibility is quite different for the two
isotopes: ${\mathcal{V}}(\mathcal{I}=0) \approx 75$\% for $^7$Li
and ${\mathcal{V}}(\mathcal{I}=0) \approx 48$\% for $^6$Li. The
best visibility achieved with lithium $^7$Li is ${\mathcal{V}}
\approx 84.5$\% \cite{miffre05b}, mostly limited by phase noise
due to vibrations \cite{jacquey06}, and the present value is less
good, because of small misalignments. The smaller visibility with
$^6$Li is due to stray $^7$Li atoms arriving on the detector after
diffraction by the second and third laser standing waves. The
variations of the visibility ${\mathcal{V}}_r$ have a very
different dependence with the current $\mathcal{I}$ for the two
isotopes, an obvious consequence of the differences in the number
of sub-levels with a given $M_F$ value and in the Land\'e factors.
We have fitted these results using equations (\ref{n1}) and
(\ref{n2}) with only two adjustable parameters, namely the
distance of the coil center to the atomic beams and the parallel
speed ratio $S_{\|}$ appearing in equation (\ref{n5}). The
agreement with the experimental data is good, the discrepancy
appearing mostly in the case of $^6$Li, when the visibility is
very small.

The fits of figure \ref{visibility6and7} assume that the signal
comes only from the isotope selected by the chosen laser
frequency. As $^7$Li is considerably more abundant than $^6$Li (
$92.4$\% vs $7.6$\%), this is, not surprisingly, an excellent
assumption for the dominant isotope $^7$Li, but this assumption
works well also with the less abundant isotope, $^6$Li. Assuming
that the fringe patterns of the two isotopes are always in phase,
we can estimate the contribution of $^7$Li isotope to the $^6$Li
experiment: from the fit, we deduce a contribution less than
$10$\% of the fringe signal. We have developed a full model of the
interferometer to explain this effect because a simple model, with
Gaussian laser beams described as top-hat beams, cannot explain
such a large isotopic selectivity.

\section{Test of velocity selectivity}

\begin{figure}
\begin{center}
\begin{tabular}{cc}
\includegraphics[width =7 cm,height= 6 cm]{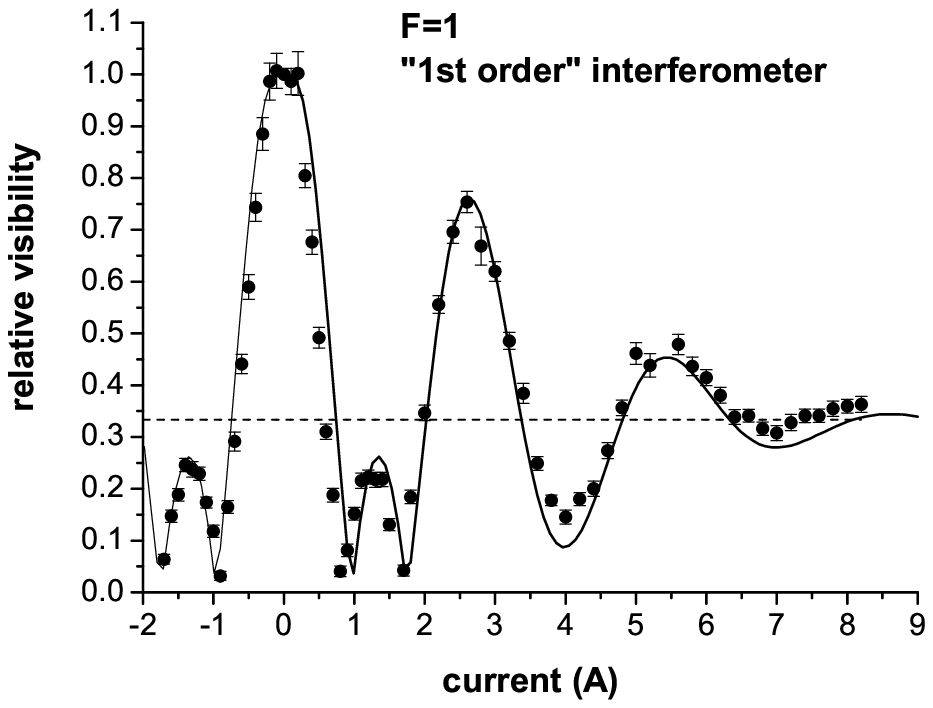}
&\includegraphics[width =7cm,height= 6 cm]{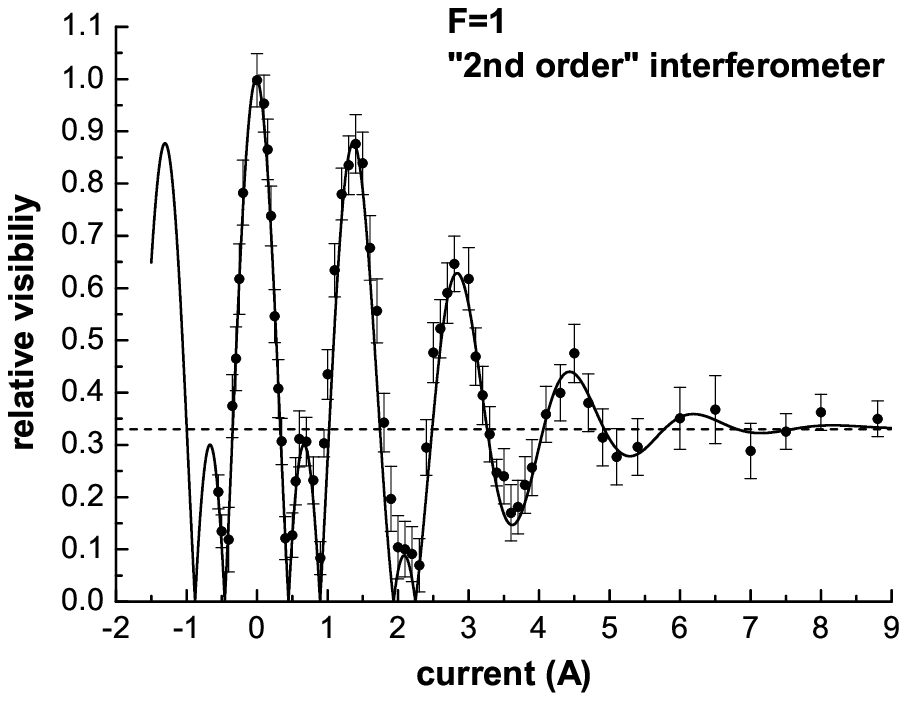}
\end{tabular}
\caption{\label{visibilityorder1and2} Relative visibility
${\mathcal{V}}_r$ (dots) of the interference fringes for $^7$Li,
pumped in its $F=1$ ground state, as a function of the coil
current. Experimental data points are represented by dots and the
fits by full curves. Left panel: first order diffraction $p=1$,
with a rapid decay of the revival intensity; right panel: second
order diffraction $p=2$, with more apparent revivals. The
asymptotic ${\mathcal{V}}_r = 1/3$ value is represented by a
dashed line.}
\end{center}
\end{figure}

For this test, we have optically pumped $^7$Li in its $F=1$ state,
using a diode laser tuned on the $^2$S$_{1/2}$, $F=2$ -
$^2$P$_{3/2}$ transition. Optical pumping must be performed before
collimation of the atomic beam, because the photon momentum
transfers due to absorptions and emissions of photons would spoil
the necessary sub-recoil collimation. In the analysis, we assume
that the three $M_F$ sub-levels of the $F=1$ states are equally
populated. We have recorded the fringe visibility using
successively the diffraction orders $p=1$ and $p=2$, with
different adjustments of the laser standing waves (beam diameters
power density, frequency detuning and mirror directions, see ref.
\cite{miffre05b}). The measured relative visibility
${\mathcal{V}}_r( \mathcal{I})$ is plotted  as a function of the
coil current $\mathcal{I}$ in figure \ref{visibilityorder1and2}:
the variations are very different from those observed on $^7$Li
without optical pumping (see figure \ref{visibility6and7}),
because now only two $|M_F|=1$ sub-levels and one $M_F=0$
sub-level are populated. When the magnetic field gradient is
large, $M_F\neq 0$ sub-levels experience a large phase shift so
that their contribution to the fringe signal is washed out by the
velocity average and the remaining fringe visibility is solely due
to the $M_F=0$ sub-level. We thus predict that ${\mathcal{V}}_r$
tends toward $1/3$ in this case because there is one $M_F=0$
sub-level over the three sub-levels of $F=1$.

As discussed above, the parameter which governs the decay of the
revivals is the parallel speed ratio and a fit of these data gives
$S_{\|} = 9.0$ when using the first order diffraction, $p=1$, and
$S_{\|} = 14.5$ when using second order diffraction, $p=2$. The
beam source conditions \cite{miffre05b} were the same in both
cases and, from our study of the lithium beam
\cite{miffre04,miffre05}, we know the initial value of the
parallel speed ratio, $S_{\|i} \approx 8.5$ . The velocity
selective character of Bragg diffraction appears to be strong for
second order diffraction.

\section{Conclusions}

In this letter, we have studied the effects of a magnetic field
gradient on the signals of a lithium atom interferometer and we
have analyzed the resulting variations of the fringe visibility.
Following Siu Au Lee and co-workers \cite{giltner}, we use a coil
to produce the magnetic field gradient rather than a septum
carrying an electric current and inserted between the two atomic
beams as done by D. Pritchard and co-workers
\cite{schmiedmayer94,schmiedmayer97}: the coil does not require
the fine alignment of the septum and the two arrangements appear
to give very similar effects.

The idea that such an experiment can measure the relative width of
the velocity distribution was pointed out by J. Schmiedmayer et
al. \cite{schmiedmayer97}. We have applied this idea with our
laser diffraction atom interferometer and we have observed a
modification of the velocity distribution due to Bragg diffraction
by comparing first and second diffraction orders. We have shown
that the visibility variations give access to other quantities,
such as the interferometer isotopic selectivity, which is
excellent in our experiment with a correct choice of laser
detuning. Finally, optical pumping modifies strongly the
visibility variations, in good agreement with simple arguments.
The ability to test the velocity distribution or the isotopic
selectivity will be very useful for the following reasons:

- as discussed after equation (\ref{n6}), the velocity
distribution of the atoms contributing to the interference signals
differs from the velocity distribution of the incident atomic beam
and this difference is very important for accurate phase shift
measurements because most phase shifts are dispersive
(proportional to $v^n$ with $n=-1$ as in a measurement of an
electric polarizability \cite{miffre06} or $n=-2$ as in the
present experiments).

-  a test of the isotopic selectivity distribution could also be
useful to measure the isotopic dependence of some quantity, for
instance the electric polarizability. This possibility has
presently little interest because this dependence is considerably
smaller than the present accuracy \cite{miffre06} but it might not
be always so.

\acknowledgments

We have received the support of CNRS MIPPU, of ANR and of R\'egion
Midi Pyr\'en\'ees.


\end{document}